\documentclass{aa}  
\usepackage{natbib}
\usepackage{comment}
\usepackage{graphicx}
\UseRawInputEncoding
\usepackage{txfonts}
\usepackage{hyperref}
\hypersetup{
    colorlinks=true,
    linkcolor=blue,
    filecolor=magenta,
    citecolor=blue,
    urlcolor=blue,
}
\begin{document}

   \title{Empirical determination of atomic line parameters of the 1.5 $\mu$m spectral region}
   \titlerunning{Determination of atomic line parameters of the 1.5 $\mu$m spectral region}
   \authorrunning{Trelles Arjona et al.}

   \author{J. C. Trelles Arjona
          \inst{1,2}
          , B. Ruiz Cobo 
          \inst{1,2}
          \and
          M. J. Mart\'inez Gonz\'alez\inst{1,2}
          }

   \institute{Instituto de Astrof\'isica de Canarias (IAC),
              ,V\'ia L\'actea s/n, 38205 San Crist\'obal de La Laguna, Tenerife, Spain
         \and
             Dept. Astrof\'isica, Universidad de La Laguna, 38205 San Crist\'obal de La Laguna, Tenerife, Spain \\
             \email{jtrelles@iac.es, brc@iac.es and marian@iac.es}
             }

 
  \abstract
   {Both the quality and amount of astrophysical data are steadily increasing over time owing to the improvement of telescopes 
and their instruments. This requires  corresponding evolution of the techniques used for obtaining and analyzing the resulting data. The 
infrared spectral range at 1.56 $\mu$m usually observed by the GRegor Infrared Spectrograph (GRIS) at the GREGOR solar telescope has a width of around 30 $\AA$ and 
includes at least 15 spectral lines. Normally, only a handful of spectral lines (five at most) are used in studies using GRIS because of the lack of atomic
 parameters for the others. Including more spectral lines may alleviate some of the known ambiguities between solar 
atmospheric parameters.}
   {We used high-precision spectropolarimetric data for the quiet Sun at 1.56 $\mu$m observed with GRIS on the GREGOR along with the 
SIR inversion code in order to obtain accurate atomic parameters for 15 spectral lines in this spectral range.}
   {We used inversion techniques to infer both solar atmospheric models and the atomic parameters of 
spectral lines which, under the local thermodynamic equilibrium (LTE) approximation, reproduce spectropolarimetric observations.}
   {We present accurate atomic parameters for 15 spectral lines within the spectral range from 15644 $\AA$ to 15674 $\AA$.
This spectral range is commonly used in solar studies because it enables the study of the low photosphere. Moreover, the infrared 
spectral lines are better tracers of the magnetic fields than the optical ones.}
 {}

   \keywords{atomic data --
                oscillator strengths --
                line: profiles --
                Sun: atmosphere --
                methods: observational
               }

   \maketitle
%

\section{Introduction}

Light is the most powerful tool for getting information from astronomical bodies. Physical conditions of the emitting 
object are encoded in the light emitted. Using spectroscopy, for example, it is possible to identify the elements present 
in stellar atmospheres simply by looking for their signatures in the spectra. These signatures are the spectral lines 
formed by transitions between the energy states of an atom.

Many efforts have been made to characterize spectral lines. In the gathering of information about spectral 
lines, it is inescapable to cite the work of \citet{moore66} and subsequent databases such as NIST \citep{nist}, VALD \citep{piskunov95}, and CHIANTI \citep{dere97}. One of the key points in these studies 
is the oscillator strength of the spectral line $f$, which can be defined as the ratio between the probability of 
absorption or emission of electromagnetic radiation in a transition between energy levels of an atom or molecule and the 
probability of this transition in a classical oscillator \citep{mihalas78}.Two important international collaborations have 
been created in order to calculate and collect oscillator strengths of astrophysical interest: the Opacity Project \citep{seaton94} and the FERRUM project \citep{johansson02}, specializing in iron.

In general, there are three ways of calculating oscillator strengths: theoretical, semi-empirical, and observational. 
A good description of the theoretical and semi-empirical methods can be found in \citet{borrero1} and references therein.

Observationally, it is possible to determine the oscillator strengths using solar spectra. The basic idea is to use 
a model atmosphere and solve the radiative transfer equation (RTE), allowing the oscillator strength value to be a free parameter 
in order to make the synthetic spectra agree with solar spectrum. It is important to stress that the results achieved using 
these studies rely on having suitable atmospheric models, line damping parameters, solar abundance elements, and so on.

The first successful attempt at determining oscillator strengths from solar spectra was by \citet{gurtovenko81,gurtovenko82}, who used the model atmosphere of \citet{holweger74} to fit the central intensity of 865 Fe I 
spectral lines (between 4000\,\AA\ and 8000\,\AA) to the solar atlas \citep{delbouille73}. The same atlas was used by \citet{thevenin89,thevenin90} to determine the oscillator strength of thousands of spectral lines from 4006 to 7950\,\AA. These latter authors 
used the model atmosphere of \citet{gustafsson75} in order to fit either the central core and the beginning of the wings 
or the central depth of the line. Those works compared some of their results with theoretical calculations \citep{blackwell1,blackwell2} and revealed a high degree of agreement.

One step further was taken by \citet{borrero1}, who introduced such improvements as fitting the whole 
spectral line, as opposed to fitting some points of the line; using broadening collisional parameters \citep{anstee2,barklem1,barklem2} where previous works had used the Uns\"old formula with correction 
factors; and using a two-component model atmosphere to simulate granular and intergranular behavior \citep{borrero2}. \citet{socasnavarro11} used the same procedure but with a model atmosphere of a single component to calculate
 the oscillator strength of the Fe I 6302.5\,\AA\ spectral line used by the solar optical telescope on the Hinode satellite \citep{kosugi07,ichimoto08,shimizu08,suematsu08,tsuneta08}. This oscillator strength value was used by \citet{socasnavarro15} to determine the oxygen abundance in the solar photosphere.

In this paper, we wish to make our contribution to this effort by observationally determining accurate parameters for 15 spectral lines of the 1.56 $\mu$m spectral region (Fig. 1). In this spectral region, the main contributor to the solar 
continuum opacity, the H- ion, reaches the minimum of opacity. It therefore allows us to probe the deepest layers of the solar 
photosphere \citep{gray08}. Furthermore, infrared spectral lines are better tracers of magnetic fields than optical ones 
because the ratio between the Zeeman splitting and the width of the line is proportional to the wavelength \citep{bellot3}.
 
Because of the high inhomogeneity of the solar photosphere and the nonlinearity of the RTE, 
there is no single model atmosphere capable of explaining an averaged spectrum over a large area of the solar disk. This 
may lead to inaccuracies when calculating oscillator strengths. Therefore, our approach to solve the problem is based on 
using not only a single solar spectrum but also spatially resolved pixel-by-pixel observations  
($0.135^{\prime\prime}$ $\times$ $0.135^{\prime\prime}$). 
Thus, we have a large number of values for the oscillator strength from which we can calculate the mean and standard 
deviation for each spectral line.

We use spectropolarimetric inversions for this work. In solar physics, many computer codes have been developed 
for spectropolarimetric inversions (e.g.,  HAZEL: \citet{andres}, SPINOR: \citet{frutiger98}, SIR: \citet{SIR}, and NICOLE: \citet{socasnavarro00}). These codes try to 
find reliable physical quantities for the solar atmosphere from a set of observed Stokes parameters. To this end, the difference 
between observed and synthetic spectra is minimized. The synthetic spectrum is numerically calculated from a simplified
 model atmosphere by solving the RTE.

\begin{figure}
  \resizebox{\hsize}{!}{\includegraphics{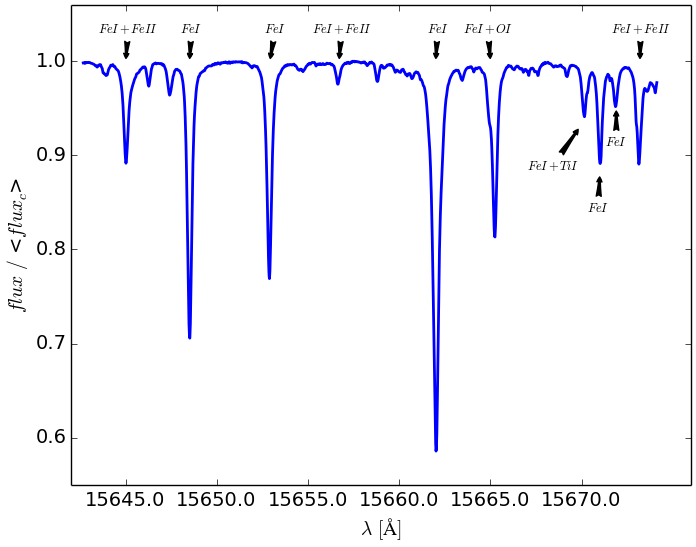}}
  \caption{Region of the FTS atlas of \citet{wallace11}. The spectral lines marked with arrows have been included in this study.}
  \label{fig:uno}
\end{figure}

\begin{figure}
  \resizebox{\hsize}{!}{\includegraphics{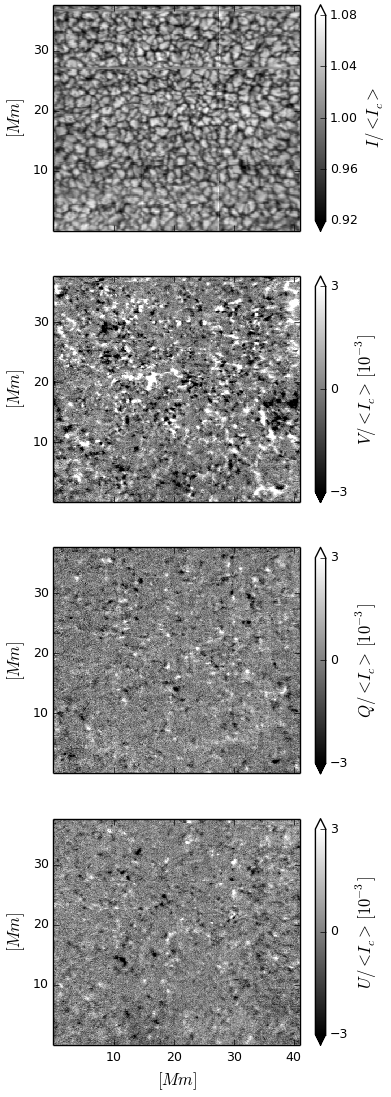}}
  \caption{Intensity and polarization maps of dataset 1 observed on 2018 August 29.}
  \label{fig:dos}
\end{figure}

\section{Observations and data reduction}

On 2018 August 29, December 14, and December 17,  the four Stokes parameters of the spectral lines at 1.56 $\mu$m (from 15644\,\AA\ 
to 15674\,\AA) were recorded using the Tenerife Infrared Polarimeter working with GRIS \citep{tip2,collados12} attached to the German GREGOR telescope \citep{schmidt12}. The data sets were obtained in a quiet 
Sun region located at disk center.
The spectral sampling was 40.00 m\AA. A spatial resolution of $0.5^{\prime\prime}$ was reached through the 
adaptive optics system \citep{berkefeld} locked on granulation.

The maps we present here (three, one per day mentioned above) cover a scanned area of about $62.0^{\prime\prime}$ $\times$ $54.0^{\prime\prime}$ 
with an integration 
time of 2 s per slit position (the total cadence was $\sim$ 3.6 s per slit position due to overheads). The slit position was aligned with the
solar north--south direction and the Sun was scanned in a perpendicular direction to the slit with a step size of $0.135^{\prime\prime}$. The
 contrast of the continuum image -- defined as the standard deviation of the full map divided by its average value -- is 2.7 per cent.

The data were demodulated with dedicated software \citep{Schlichenmaier02}. This software also corrects for bias, flatfield, and bad pixels and removes the instrumental cross-talk. Additional corrections were applied to the data:

-We corrected for wavelength-independent stray light by subtracting a veil quantity ($I_{\rm v}$) from  our intensity spectrum:
\begin{equation}
    I_{\rm corr} = \frac{I - I_{\rm v}}{I_{\rm c} - I_{\rm v}}\,,
\end{equation}
where $I_{\rm c}$ stands for the continuum intensity and $I_{\rm corr}$ for the veil-corrected spectrum. In order to infer the veil, we fit the 
average intensity spectrum to that of the Fourier
Transform Spectrometer (FTS) of \citet{wallace11}. As the FTS spectrum has higher spectral resolution, 
we convolved it with a Gaussian function with a standard deviation $\sigma$ that will be representative of our spectral resolution. We 
then inferred the value of both $\sigma$ and the veil by minimizing the quadratic difference between our average Stokes $I$ and the
 convolved atlas. The $\sigma$ was 81.26 m\AA\ and the veil was 2.5\%. We used these numbers to correct each Stokes $I$ profile. 

After the standard reduction, the continuum is not flat but has some undulations. We fit a high-order polynomial to the continuum 
wavelengths and divide by it. All Stokes profiles were normalized accordingly.

-We removed the residual cross-talk from Stokes $I$ from Stokes $Q$, $U,$ and $V$. We calculated the mean value in the continuum spectral 
region for 
each pixel and the polarization profiles. In the Zeeman effect, there is no continuum polarization, and we do not expect the continuum to be 
polarized by scattering at such  a long wavelength. Assuming that the polarized continuum should be zero, we then calculated the correction 
factor $\delta$ for the cross-talk from intensity as follows: 

\begin{equation}
    \delta = \frac{\sum_i S (\lambda_{i})}{\sum_i I(\lambda_{i})}\,,
\end{equation}

\noindent where $S$ runs the $Q$, $U,$ and $V$ polarization profiles,
  and the sub-index $i$ contains the selected continuum wavelengths. $S (\lambda_{\rm c})$ and 
$I(\lambda_{\rm c})$ represent mean values in the continuum region.
Finally, the cross-talk correction is:

\begin{equation}
    S_{\rm corr}^{(\lambda)} = S(\lambda) + \delta I(\lambda)\,,
\end{equation}

-We denoised the data using a principal component analysis (PCA; \citet{loeve55}; \citet{rees03}). This technique is used to reduce 
the uncorrelated noise but not the systematic errors. It works by decomposing the data set into its principal components or eigenvectors and 
sorting them 
by their eigenvalues. The data can then be rebuilt using just the eigenvectors with highest eigenvalues 
(see \citet{marian2008a,marian2008c} for more details).

-We corrected for polarized interference fringes. The fringing of charge coupled device (CCD) detectors occurs because of interference between the incident light 
and the light internally reflected at the interfaces between the thin layers of the CCD. Fringes can be observed easily after PCA denoising, 
and they change with time and slit position. We removed them by fitting, pixel by pixel, a sinusoidal function to the blue and red continuum 
close to each spectral line. These two sinusoidal functions usually have different frequencies or amplitudes, and so we linearly interpolate from one 
function to the other in order to discover how fringes affect each spectral line region. In Fig.\ 3 we show an example of a Stokes $Q$ map 
at time-step 300, before (left) and after (right) fringe removal. 

\begin{figure}
  \resizebox{\hsize}{!}{\includegraphics{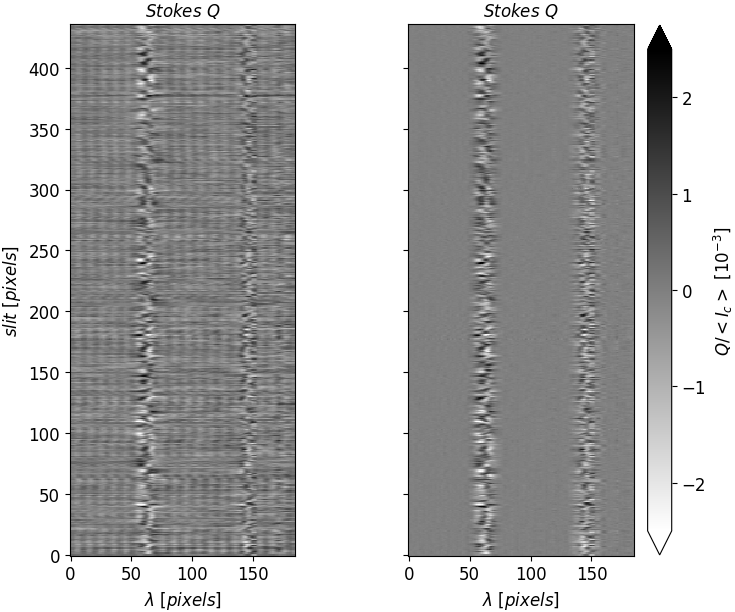}}
  \caption{Stokes $Q$ map for time-step number 300 after applying PCA denoising, before (left) and after (right) fringe removal.}
  \label{fig:tres}
\end{figure}

\section{Determination of atomic parameters}

Historically, the spectral lines in the visible spectral range have been studied in depth in the laboratory, but experimental data are lacking for the 
infrared part of the spectrum. When there are no accurate laboratory measurements of atomic parameters for certain transitions, such as the 
oscillator strength, the solar spectrum can be used to infer them. Previous works fixed the model atmosphere and performed an inversion of the 
observed spectrum with the oscillator strength as the only free parameter \citep{gurtovenko81,gurtovenko82,thevenin89,thevenin90,borrero1}. The reason for fixing the model atmosphere is that the oscillator strength and certain atmospheric parameters, 
mainly the temperature, 
are degenerate. For instance, the temperature is degenerate with the oscillator strength because it influences the strength of the spectral 
line; temperature determines the number density of neutral iron atoms, the number density of the atoms excited to the lower 
level of the transition, the gradient of the source function, and so on.

In this work, we propose an iterative method in which the model atmosphere and the oscillator strength are inferred simultaneously. 
In one iteration we fix the model atmosphere, in the next we fix the oscillator strength, and so on. Thus, we search for the best possible match between
 the synthesis and the observation.
Furthermore, we apply this to a large number of spectra emerging from 1D atmospheres corresponding to the individual pixels. Hence, 
statistically this method should be more robust than in previous works.

The SIR (Stokes Inversion based on Response Functions) code \citep{SIR} was used to invert our spectropolarimetric 
data. This code allows us to synthesize and invert spectral lines under the assumption of LTE (when collisional 
transitions dominate over radiative ones, e.g.,\ in solar photospheric layers) by solving the RTE for
 polarized light:

\begin{equation}
    \frac{d\textbf{I}}{d\tau_{\rm c}} = \textbf{K}(\textbf{I}-\textbf{S})
,\end{equation}
where $\tau_{c}$ is the optical depth at the continuum wavelength, \textbf{I} and \textbf{S} are the pseudo-vectors of the Stokes parameters 
and the source function, respectively, and \textbf{K} stands for the propagation matrix defined as

\begin{equation}
    \textbf{K} = \frac{1}{\kappa_{{\rm c}_{\tau}}}(\kappa_{\rm c}\textbf{1} + \kappa_{\rm l}\textbf{$\Phi$})
,\end{equation}
where \textbf{1} and \textbf{$\Phi$} are the identity and line absorption matrixes, 
$\kappa_{\rm c}$ and $\kappa_{{\rm c}_{\tau}}$ stand for the continuum opacity and the 
continuum opacity evaluated at the reference wavelength, and $\kappa_{\rm l}$ is the line opacity given by

\begin{equation}
    \kappa_{\rm l} =  \kappa_{\rm l}^{A} - \kappa_{\rm l}^{S}
,\end{equation}
where $\kappa_{\rm l}^{A}$ is the line absorption coefficient and $\kappa_{\rm l}^{S}$ stands for the stimulated emission coefficient. 
Through the comparison of the RTE derived from quantum mechanics and derived from the classical 
physics (more details in \citet{landi04}, the line absorption coefficient can be written in the form

\begin{equation}
    \kappa_{\rm l}^{A} = \frac{\pi e_{0}^{2}}{mc}N_{\rm l}f(\alpha_{\rm l}J_{\rm l} \rightarrow \alpha_{\rm u}J_{\rm u})
,\end{equation}
where $e_{0}$ is the absolute value of the electron charge, $N_{\rm l}$ is the number of atoms per unit volume in the lower level of the 
transition, $m$ and $c$ are the electron mass and the speed of light, and $f(\alpha_{\rm l}J_{\rm l} \rightarrow \alpha_{\rm u}J_{\rm u})$ 
is the oscillator 
strength of the transition.

The strength of a spectral line, defined as the area under or above (i.e.,\ emission or absorption) the curve described for the spectral 
line in a wavelength versus intensity plot, depends on the product $Agf$, where  $A$ is the abundance (related to $N_{\rm l}$ in Eq.\ 7). 
The oscillator strength $f$ is the correction factor when we express the probability of absorption of electromagnetic radiation in 
terms of the degeneracy of the lower level $g$ and the transition probability of a harmonic oscillator \citep{mihalas78}. As can be seen 
in Eqs. 4 to 7, both the abundance and oscillator strength affect the RTE. Consequently, accuracy in the 
determination of these quantities is  key to getting accurate results in the inversions.

The parameter commonly used for the synthesis of spectral lines is the logarithm of the product of the lower level degeneracy multiplied by 
the oscillator strength $log(gf)$. The abundance has to be fixed in order to calculate the oscillator strength. We therefore take 
the abundances from \citet{asplund}.

We use the coefficients for collisional broadening from the ABO theory \citep{anstee1,anstee2,barklem1,barklem2}, where $\sigma$ and $\alpha$ stand for the line broadening cross section and velocity parameter, 
respectively. The coefficients presented in Table 1 were calculated by Barklem (private communication).

\subsection{Models}

SIR uses an hermitian method to integrate the RTE for polarized light \citep{bellot1}. The great 
accuracy and speed reached by an hermitian method is shown in \citet{delacruz13}, although, following their 
calculations, a method based on Diagonal Element
Lambda Operator (DELO) with cubic Bezier interpolation of the source function can in some cases reach a better accuracy, mainly in 
chromospheric layers. Nevertheless, the accuracy gain depends on the smoothness of the different physical 
quantities with depth, and for photospheric lines an hermitian method behaves even better
than a cubic Bezier method. For an in-depth analysis of the performance of different integration schemes see
\citet{janett17} and \citet{janett18}. \citet{bellot1} calculate an accuracy better than $10^{-5}$ for integration 
of Stokes $I$ and $10^{-6}$ the other Stokes profiles when a step size of 0.1 in the logarithm of continuum optical depth is chosen. 
Consequently, in this paper we have used models with 55 points in depth, from 1.4 to $-$4.0 in $\log(\tau)$ with a step size of 0.1.
As a boundary condition for the integration of the RTE, SIR uses a diffusion approximation; that is, at the bottom of the atmosphere, 
the first point in the grid model, the Stokes vector is approximated by:

\begin{equation}
   I_\nu = S_\nu+ K^{-1}\frac{dS_\nu}{d\tau_5}
,\end{equation}
where $S_\nu=(B_\nu,0,0,0)$.

The models have three quantities that are independent of depth, namely macroturbulence velocity, filling factor, and stray light 
contamination (not used in this work), and also 11 depth-dependent quantities, namely the logarithm of line-of-sight 
continuum optical depth at 5000\,\AA, temperature ($T$), electron pressure ($P_{\rm e}$), microturbulence velocity, magnetic field 
(strength, inclination, and azimuth), line-of-sight velocity, geometrical height ($z$), gas density ($\rho$), and gas pressure 
($P_{\rm g}$). Not all these quantities are completely independent:  $z$, $\rho$, and $P_{\rm g}$ are evaluated from $T$ and $P_{\rm e}$. 
Additionally, although SIR can allow the determination of $P_{\rm e}$ as a free parameter, we have chosen the hydrostatic equilibrium 
option: $P_e$ is calculated from $T$ and abundances assuming hydrostatic equilibrium through the ideal gas law assuming a $P_g$ 
value at the upper boundary of the atmosphere. 
In order to evaluate the electron number density, SIR uses Saha's equation for 28 elements, H$^-$ , and H and He molecules. As in this 
equation the ion populations depend on electron density, SIR uses an iterative procedure based on \citet{mihalas67}. A detailed 
description of the state equation and the algorithm used to determine partial pressures, damping parameters, and continuum 
absorption coefficient can be found in \citet{wittmann74}.

\subsection{Estimation of $\log(gf)$ values}

As the initial step, for each dataset we inferred $\log(gf)$ of the 15662.017\,\AA\ line from its intensity profile averaged 
over the field of view. This line was selected as it is the strongest. The model atmosphere was fixed to 
 model C of \citet{fontenla93} (FALC), which is a semi-empirical stratified model tuned to reproduce the solar 
quiet spectrum. After several inversion runs looking for the most suitable macroturbulence velocity, a fixed value of 0.92 
km s$^{-1}$ was found to be optimum. 
 From these three inversions, an average value of $\log(gf)=0.140 \pm 0.018$ was obtained.

Subsequently, we performed inversions of these intensity-averaged profiles (three, one per data set) using the FALC model as 
an initial atmospheric model but fixing $\log(gf)$ to the previously obtained value. The stratifications of the temperature and 
line-of-sight velocity of the model were modified during the inversion. Through the inversion, a more suitable model is obtained
 which is used to recalculate the  $\log(gf)$ value for this spectral line.

\begin{figure}
  \resizebox{\hsize}{!}{\includegraphics{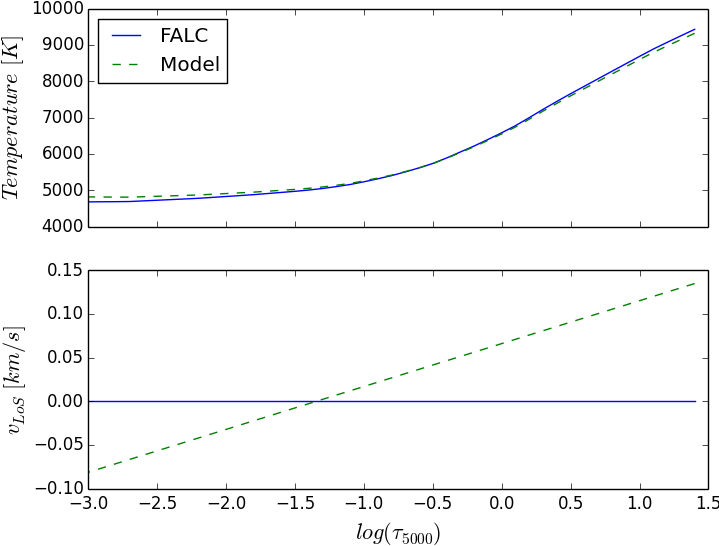}}
  \caption{Comparison between FALC and the new model in temperature (top) and line-of-sight velocity (bottom). The solid blue line 
represents the FALC model. The dashed green line represents the new model.}
  \label{fig:cuatro}
\end{figure}
 
This whole process was repeated iteratively to find the value of $\log(gf)$ and the atmospheric model that minimize the difference 
between the observed and synthetic spectrum. 

Figure\ 4 shows the temperature and line-of-sight velocity of FALC and the new model for comparison. The temperature of 
both models is very similar. However, a linear gradient along the optical depth of the line-of-sight velocity is needed to fit the 
observations. This gradient is the result of the known asymmetry produced in the presence of granules and intergranules in the averaged spectra.

Using this new model, the oscillator strength of two more spectral lines (at 15648.514 and 15652.873 \AA) could be calculated using 
$\log(gf)$ as a free parameter in the synthesis. Figure 5 shows the fits between the observation and the synthesis for the three 
spectral lines. The mean and the standard deviation of the $\log(gf)$ values for the three measurements and for each line are: 
$-0.596 \pm 0.007$ for the line at 15648.514 \AA, $-0.025 \pm 0.003$ for the line at 15652.873 \AA,~ and $0.187 \pm 0.012$ for the 
line at 15662.017 \AA. Inversion uncertainties in the determination of $log(gf)$ values were calculated for the line at 
15662.017 \AA~ resulting in 0.010. Thus, the inversion uncertainties (i.e.,\ the sensitivity of the spectral line to changes in the
 $\log(gf)$ values) are contained in the errors of the $\log(gf)$ calculation we present here.
As can be seen in Fig.\ 5, the observed profiles are well reproduced for the synthesis.

\begin{figure}
  \resizebox{\hsize}{!}{\includegraphics{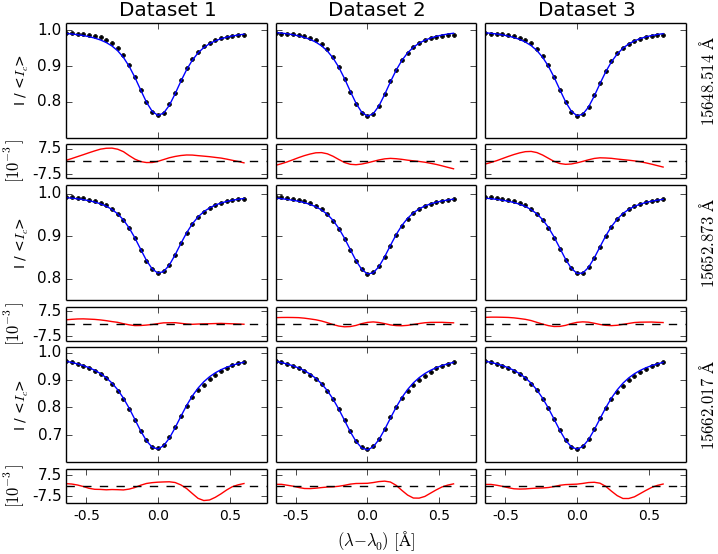}}
  \caption{Fits between the observation (black points) and the synthesis (blue solid lines) of each data set and each spectral 
line. The residuals are shown as a solid red line.}
  \label{fig:cinco}
\end{figure}

At this point we faced the problem of the degeneracy between temperature and $\log(gf)$ values. As explained above, the high 
inhomogeneity of the solar photosphere and the nonlinearity of the RTE may lead to inaccuracies in 
model atmosphere determination. Changes in temperature affect the number of absorbents that produce the spectral lines ($N_{\rm l}$ in 
Eq. 7). Furthermore, the source function ($S$ in Eq. 4) in LTE conditions is the Planck function, which is related to the
 temperature. The effects produced by temperature may be compensated for by changing the $\log(gf)$ values in order to fit the 
observation (e.g. as seen in Eq. 7). The difference in temperature at the solar photosphere between granules 
and intergranular lanes could reach around 1000~K \citep{basilio96}, but fluctuations in temperature will be smaller 
inside each pixel on the plane of the sky. We therefore used the highest-quality data set (the one recorded on 2018 
August 29) for doing inversions pixel by pixel to overcome this problem. Thus, we obtained $\log(gf)$ values with different 
temperatures and we were able to calculate the mean value and standard deviation.

To this end, a new inversion was done using the three spectral lines with their now known $\log(gf)$ values (hereinafter, main spectral 
lines). The strategy followed to do the inversions was the same as before (explained in Appendix B). Hereafter, the model 
used to perform the fit consists of two 1D atmospheres (magnetized and nonmagnetized). It is necessary to include the magnetized atmosphere
 because the magnetic field affects abundances determination \citep{borrero3}. Moreover, we included the Stokes $Q$, $U,$ and $V$ parameters
for the 15648.514\,\AA\ spectral line (i.e.,\ the spectral line with highest magnetic sensitivity) in the inversions to provide 
more information. Thus, $\log(gf)$ values are more constrained. The 400 pixels with the best $\chi^{2}$ between the observation 
and the synthesis were chosen from the inversion. Two atmospheric
models were obtained for each pixel. These models were used 
to calculate $\log(gf)$ for each line within the GRIS spectra around 1.56 $\mu$ pixel by pixel. Thus, it was possible to 
redetermine the $\log(gf)$ values of the main spectral lines and get a first approximation for the $\log(gf)$ values for the 
remaining spectral lines.

\begin{figure}
  \resizebox{\hsize}{!}{\includegraphics{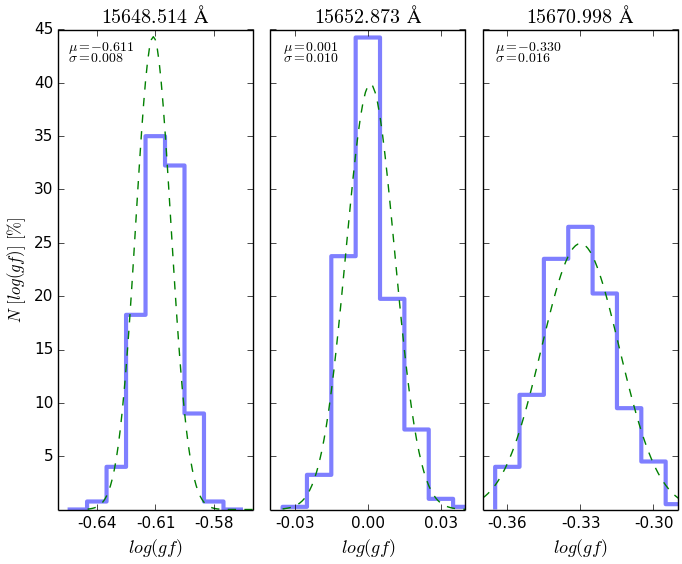}}
  \caption{$\log(gf)$ histograms (blue solid line) for three spectral lines (from left to right, 15648.514 \AA, 15652.873 \AA,~ 
and 15670.998 \AA). The dashed green line is a Gaussian function with the mean value and standard deviation of $\log(gf)$.}
  \label{fig:seis}
\end{figure}

Finally, the $\log(gf)$ values of the 15 spectral lines obtained in the previous step were used to perform a new inversion. 
The models retrieved in this inversion were used to redetermine the $\log(gf)$ values of all the spectral lines.

Figure 6 shows the histograms of the $\log(gf)$ values for three spectral lines. The plots were made using the 400 best 
pixels with the best $\chi^{2}$ between the synthesis and the observation, as noted before. In order to show the adequacy of 
the method, a Gaussian function is plotted over the histogram with the mean and standard deviation values calculated for 
the $\log(gf)$ value for each spectral line.

\subsection{Dealing with blends}
Blends occur when two or more spectral lines are so close together that they are impossible to study separately. In this
 work we faced two types of blends: 

-Secondary line over the wing of a principal line: The $\log(gf)$ of the principal spectral line was calculated with the method 
explained above but excluding the wing contaminated by the blend. The $\log(gf)$ value of the principal spectral line was 
fixed in order to get the $\log(gf)$ value of the secondary line.

-Secondary and principal line almost sharing the central wavelength: Firstly, the range of $\log(gf)$ of the principal and secondary 
lines were calculated separately. This range is delimited by two values. The upper bound of the $\log(gf)$ of the principal line is 
given by the difference in depth between the synthesized and the observed line. The lower bound occurs when the synthetic line by itself is below the noise level. Secondly, $\log(gf)$ values of the 
principal line were fixed and the $\log(gf)$ values of the secondary line were found using the difference between the synthesis and 
the observations ($\chi^{2}$) for a large number of pixels. Finally, the two values that minimized the $\chi^{2}$ were selected.

\subsection{Discussion of results}

The $\log(gf)$ values calculated in this work are listed in Table 1 and the whole iterative process is depicted in Fig. 8.

The results obtained for the $\log(gf)$ values are in accordance with those previously calculated in the literature. For example, 
the $\log(gf)$ values for the spectral lines at 15648.514 and 15652.873 \AA~ calculated by \citet{borrero1} are 
$-$0.675 and $-$0.043 respectively. In their work, those authors used an Fe abundance  of 7.43 dex \citep{bellot2}. However, 
we used an Fe abundance of 7.50 dex \citep{asplund}, which gives  $\log(gf)$ values of $-$0.745 and 
$-$0.113, very close to the values of this work.

It is important to stress that the values for the oscillator strengths presented in this work were calculated with the 
abundances from \citet{asplund}: 7.50, 8.69, and 4.95 dex for iron (Fe), oxygen (O), and titanium (Ti), respectively. Oscillator 
strength values are useless in observational studies without their corresponding abundances. If the abundances are changed, 
the values of the oscillator strengths must be changed proportionately.

The excellent agreement with previous works and the low values of the uncertainties found here are good indicators of the quality 
of the method. Moreover, the residuals ---defined as the difference between observed and synthetic spectra--- resulting in the last 
inversion given in Fig. 7 prove the adequacy of the models inferred.

The uncertainties are higher for the secondary spectral lines in blends (see Table 1). This was expected because they have 
a small influence on the inverted model atmosphere owing to their low significance in the conditions of temperature and electronic 
density where the spectrum is formed in the solar atmosphere.

\begin{figure}
  \resizebox{\hsize}{!}{\includegraphics{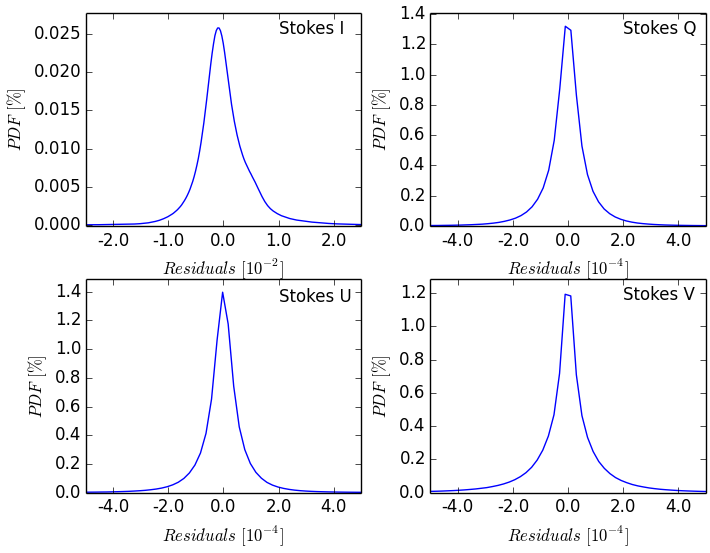}}
  \caption{Residuals of the whole map resulting from the last inversion. Upper panels: Residuals for Stokes $I$ (left) and Stokes 
$Q$ (right). Lower panels: Residuals for Stokes $U$ (left) and Stokes $V$ (right).}
  \label{fig:siete}
\end{figure}

 \begin{figure*}
\centering
   \includegraphics[width=17cm]{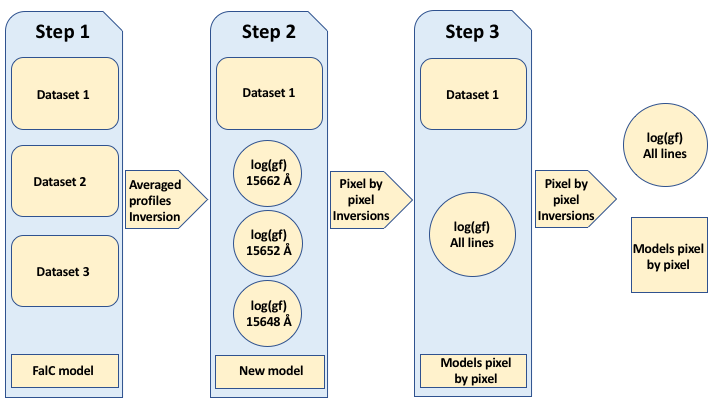}
     \caption{Scheme of the process. In step 1, three averaged profiles coming from three different data sets ---together with the FalC model--- are 
used to calculate the oscillator strengths of three spectral lines. In step 2, pixel-by-pixel inversions are done using the oscillator 
strengths of the three spectral lines calculated previously and the data set with the highest quality to determine the oscillator strengths 
of 15 spectral lines. In step 3, step 2 is repeated but taking into account 15 spectral lines in the inversions to redetermine 
the oscillator strengths of all the spectral lines.}
     \label{fig:ocho}
\end{figure*}

As mentioned above, the coefficients for collisional broadening come from the ABO theory \citep{anstee1,anstee2,barklem1,barklem2}, but it was not possible to calculate coefficients with asterisks in Table 1. In these cases, 
values of the order of known values were taken. Therefore, more accurate $\log(gf)$ values for the secondary spectral lines in blends 
could be reached with the real values of the $\sigma$ and $\alpha$ coefficients.

\begin{figure*}
\centering
   \includegraphics[width=17cm]{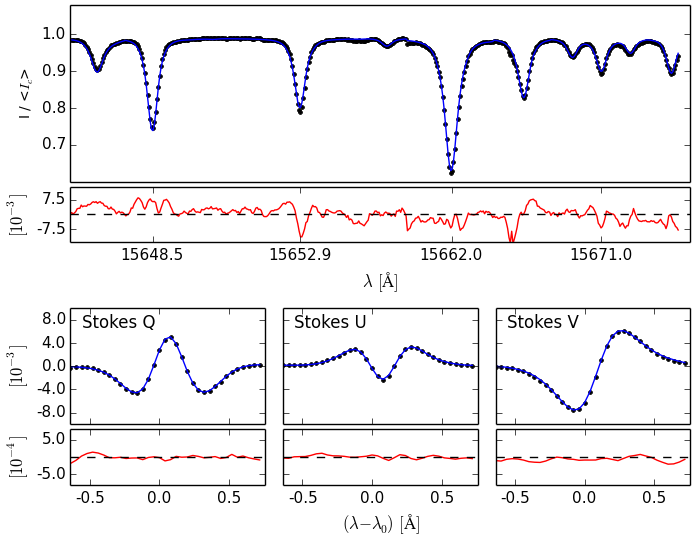}
     \caption{Fits achieved in the last inversion step between the observation (black points) and the synthesis (blue solid lines) in a 
single pixel. Upper panel: Intensity profiles. Lower panel: Stokes $Q$, $U,$ and $V$ of the 15648.514\,\AA\ spectral line. The residuals are 
shown with a red solid line.}
     \label{fig:nueve}
\end{figure*}

\begin{table*}[htb]
\caption{\label{t1} Atomic line parameters of the 1.5 $\mu$m spectral region. The columns show (from left to right) the element and the 
ionization state, wavelength, total angular momentum quantum number of the lower and upper levels ($J_{\rm l}$ and $J_{\rm u}$), 
the oscillator strength 
$\log(gf)$ calculated in this work, $\log(gf)$ values from the literature, the excitation potential of the lower level ($\chi_{\rm e}$), the Land\'e 
$g$-factor, and the coefficients for collisional broadening ($\alpha$ and $\sigma)$. Abundances: \citet{asplund} (7.50, 8.69, and 4.95 
dex for Fe, O, and Ti, respectively).}
\centering 
\begin{tabular}{c c c c c c c c c c} 

\hline\hline 
Element & Wavelength (\AA) & $J_{l}$ & $J_{u}$ & $log(gf)$ & $log(gf)$ & $\chi_{e}$ (eV) & $g_{eff}$ & $\alpha$ & $\sigma$ $(a_{0}^{2})$\\

\hline 
Fe I & 15645.020 & 2.0 & 2.0 & $-0.348 \pm 0.013$ & -0.520\tablefootmark{b} & 6.311 & $2.333^{LS}$  & 0.290 &  1883 \\

\hline 
Fe II & 15645.303 & 2.5 & 3.5 & $2.481 \pm 0.144$ & & 12.897 & $0.714^{LS}$  & 0.300* & 1071* \\

\hline 
Fe I & 15648.514 & 1.0 & 1.0 & $-0.611 \pm 0.008$ & -0.745\tablefootmark{a} & 5.426 & $3.000^{LS}$  & 0.229 &  975\\

\hline 
Fe I & 15652.873 & 5.0 & 4.0 & $0.001 \pm 0.010$ & -0.113\tablefootmark{a} & 6.246 & $1.670^{jK}$  & 0.330 & 1445\\

\hline 
Fe I & 15656.635 & 5.0 & 4.0 & $-1.534 \pm 0.044$ &  & 5.874 & $1.875^{LS}$  & 0.236 &  830\\

\hline 
Fe II & 15656.639 & 3.5 & 2.5 & $-0.096 \pm 0.326$ & & 10.500 & $1.929^{LS}$  & 0.300* & 1071*  \\

\hline 
Fe I & 15662.017 & 5.0 & 4.0 & $0.200 \pm 0.013$ & 0.120\tablefootmark{b} & 5.828 & $1.500^{LS}$  & 0.240 &  1197\\

\hline 
Fe I & 15665.245 & 1.0 & 1.0 & $-0.386\pm 0.009$ & -0.490\tablefootmark{b} & 5.979 & $0.750^{LS}$  & 0.230 &  1280\\

\hline 
O I & 15665.098 & 1.0 & 0.0 & $0.456 \pm 0.107$ & & 12.087 & $0.500^{LS}$  & 0.300* & 1071*   \\

\hline 
Fe I & 15670.131 & 5.0 & 5.0 & $-0.930 \pm 0.026$ & & 6.206 & $1.550^{LS}$ & 0.282 &  1679 \\

\hline 
Ti I & 15670.157 & 5.0 & 5.0 & $1.045 \pm 0.177$ & & 5.210 & $1.214^{jK}$  & 0.326 &  1564   \\

\hline 
Fe I & 15670.998 & 1.0 & 2.0 & $-0.330 \pm 0.016$ & & 6.329 & $0.424^{jK}$  & 0.321 &   1768 \\

\hline 
Fe I &15671.870 & 3.0 & 3.0 & $-1.165 \pm 0.030$ & & 5.921 & $1.083^{LS}$  & 0.236 &  1198 \\

\hline 
Fe I & 15673.160 & 5.0 & 6.0 & $-0.488 \pm 0.020$ & & 6.246 & $0.564^{jK}$  & 0.330 &  1424 \\

\hline 
Fe II & 15673.159 & 3.5 & 2.5 & $2.971 \pm 0.144$ &  & 13.218 & $1.500^{LS}$  & 0.300* & 1071*   \\

\end{tabular}

\tablefoot{$*$ indicates assumed coefficients for collisional broadening. $LS$ and $jK$ stand for the Land\'e $g$-factors 
calculated using LS and jK coupling schemes, respectively.\\
\tablefoottext{a}{\citet{borrero1}\\}
\tablefoottext{b}{\citet{beck}}}
\label{table:log(gf)} 
\end{table*}

Figure 9  shows an example of the fits acquired through the last inversion. The high quality of the fits between the observation 
and the synthesis can be observed.

\section{Conclusions}

We have developed a method for determining accurate $\log(gf)$ values for 15 spectral lines in the 1.56 $\mu$m spectral
 region from observed solar spectroscopic data. This spectral range is usually observed with GRIS at the GREGOR telescope. 
We used very high-quality spectropolarimetric data, the SIR inversion code,  and the coefficients for 
collisional broadening. The method is an iterative process. It starts from the FALC model and the averaged intensity 
profile of observations. It finds both model atmospheres and $\log(gf)$ values which minimize the difference between 
the synthesis and the observation.

We hope this study will be very useful for the community, not only for the parameters but for the method itself, because it 
can be easily used for other spectral lines.

\begin{acknowledgements}
The authors are especially grateful to Paul Barklem, Manuel Collados Vera, Jes\'us Plata Su\'arez, Ivan Mili\'{c} and Terry Mahoney for very interesting 
discussions and comments to improve the manuscript. We acknowledge financial support from the Spanish Ministerio de Ciencia, Innovaci\'on y Universidades through project PGC2018-102108-B-I00 and FEDER funds. JCTA acknowledges  financial support by the Instituto de Astrof\'isica de Canarias through Astrof\'isicos Residentes fellowship. MJMG acknowledges financial support through the Ram\'on y Cajal fellowship. The observations used in this study were taken with the GREGOR telescope, located at Teide observatory (Spain). 
The 1.5-metre GREGOR solar telescope was built by a German consortium under the leadership of the Leibniz-Institute for Solar 
Physics (KIS) in Freiburg with the Leibniz Institute for Astrophysics Potsdam, the Institute for Astrophysics G\"ottingen, 
and the Max Planck Institute for Solar System Research in G\"ottingen as partners, and with contributions by the Instituto de 
Astrof\'\i sica de Canarias and the Astronomical Institute of the Academy of Sciences of the Czech Republic.
This paper made use of the IAC Supercomputing facility HTCondor (http://research.cs.wisc.edu/htcondor/), partly financed by 
the Ministry of Economy and Competitiveness with FEDER funds, code IACA13-3E-2493.
\end{acknowledgements}

\bibliographystyle{aa}
\bibliography{aanda}

\section*{Appendix A: Effective Land\'e factor of the Fe I 15652.873\,\AA\ spectral line}

The pair of Fe I spectral lines at 15648.514 and 15652.873\,\AA\ has been used often since its interesting  diagnostic 
capabilities were first probed \citep{solanki92}. 
The common value in the literature for the effective Land\'e factor of the Fe I line at 15652.873\,\AA\ is 1.53 \citep{solanki92}. 
However, we recalculated all the atomic parameters of this line and found a slightly higher effective Land\'e factor of 1.67. 
Provided the lower energy level of the line has an L-S coupling scheme, we can use Eq. A9 to calculate its Land\'e factor: 
\renewcommand{\theequation}{A\arabic{equation}}

\begin{equation}
    g_{_{LS}} = 1 + \frac{1}{2} \frac{J(J+1)+S(S+1)-L(L+1)}{J(J+1)}\,,
\end{equation}

\noindent where $L$ and $S$ stand for the total orbital angular momentum and the total spin, respectively, and $J$
 is the total angular momentum.
On the other hand, the upper energy level has a jK coupling scheme, therefore we used Eq. A10 \citep{landi04} to calculate its Land\'e factor:

\begin{equation}
    g_{_{J_{1}l}} = 1 + \gamma(J,\frac{1}{2},K) + \gamma(J,K,\frac{1}{2})\gamma(K,J_{1},l)\gamma(J_{1},S_{1},L_{1})\,,
\end{equation}

\noindent where $L_{1}$ and $S_{1}$ are the orbital angular and spin momentum of the `parent' energy level. $J_{1}$ is the total 
angular momentum of that level which is coupled with the orbital momentum $l$ of a further electron to give an angular 
momentum $K$. Lastly, $K$ is coupled with the electron spin to give the total angular momentum $J$ (see \citet{landi04} for details). \newline
\newline
In Eq. A10, $\gamma$ is: 
\begin{equation}
    \gamma(A,B,C) = \frac{A(A+1)+B(B+1)-C(C+1)}{2A(A+1)}\,.
\end{equation}

Finally, we obtained the effective Land\'e factor of the transition using Eq. A12.

\begin{equation}
    g_{_{\rm eff}} = \frac{g_{1}+g_{2}}{2} + \frac{(g_{1}-g_{2})J_{1}(J_{1}+1) - J_{2}(J_{2}+1)}{4}\,,
\end{equation}
where $g_{1}$ and $g_{2}$ are the Land\'e factors of the lower and upper energy levels, respectively.

\section*{Appendix B: Inversion strategy}

The inversion strategy relied on a free node configuration, which means that the inversion code takes as many nodes as it 
needs, taking into account the data and response functions (i.e.,\  the sensitivity of each Stokes parameter to each atmospheric 
quantity). SIR uses an algorithm that determines the optimum number of nodes for each quantity as a function of the number 
of zeros of the derivative of the chi-squared (more details in \citet{deltoro16}):

\renewcommand{\theequation}{B\arabic{equation}}

\begin{equation}
    \frac{\partial\chi^{2}}{\partial x_{m}} = 
\frac{2}{N_{f}}\sum_{s=0}^{3}\sum_{i=1}^{q}[I_{s}^{\rm obs}(\lambda_{i})- I_{s}^{\rm syn}(\lambda_{i};x)]w_{s,i}^{2}R_{m,s}(\lambda_{i})\,,
\end{equation}

\noindent where $N_{f}$ stands for the number of degrees of freedom (i.e.,\ the difference between the number of observables 
and that of the free parameters), $I_{s}^{\rm obs}(\lambda_{i})$ and $I_{s}^{\rm syn}(\lambda_{i};x)$ are the observation and 
the synthesis, respectively, $w_{s,i}^{2}$ are the weights (for taking into account the errors in observations, but 
they are normally kept at unity), and $R_{m,s}(\lambda_{i})$ are the response functions. 

Fits achieved using the free node configuration are slightly better than limiting the number of nodes by hand. Moreover, 
even though we give such freedom to the code, the number of nodes selected by SIR is low (normally between 1 and 5
 nodes for each quantity) and the resulting atmospheric models are smooth.
The filling factor $\alpha$ is also inverted to evaluate the fraction of the resolution element occupied by the magnetic atmosphere.

\end{document}